\documentclass[prb,preprint,amsmath,amssymb,superscriptaddress,notitlepage]{revtex4-1}
\usepackage{epsfig}
\usepackage{graphicx}
\usepackage{color}
\usepackage{bm}
\usepackage{upgreek}

\begin{document}

\vspace*{-1cm}

\title{Macroscopic time reversal symmetry breaking by \\ staggered spin-momentum interaction  
}

\author{Helena~Reichlova}
\thanks{These two authors contributed equally}
\affiliation{Institut f{\"u}r Festk{\"o}rper- und Materialphysik and W{\"u}rzburg-Dresden Cluster of Excellence ct.qmat, Technische Universit{\"a}t Dresden, 01062 Dresden, Germany}
\affiliation{Institute of Physics, Czech Academy of Sciences, Cukrovarnick\'a 10, 162 00, Praha 6, Czech Republic}
\author{Rafael Lopes Seeger}
\thanks{These two authors contributed equally}
\affiliation{Univ. Grenoble Alpes, CNRS, CEA, Grenoble INP, Spintec, F-38000 Grenoble, France}
\author{Rafael Gonz\'{a}lez-Hern\'{a}ndez} 
\affiliation{Grupo de Investigaci\'{o}n en F\'{\i}sica Aplicada, Departamento de F\'{i}sica,Universidad del Norte, Barranquilla, Colombia}
\affiliation{Institut f\"ur Physik, Johannes Gutenberg Universit\"at Mainz, 55128 Mainz, Germany}
\author{Ismaila Kounta}
\affiliation{CINaM - Campus de Luminy case 913 - 163 Avenue de Luminy - 13288 Marseille cedex 9, France}
\author{Richard~Schlitz}
\affiliation{Institut f{\"u}r Festk{\"o}rper- und Materialphysik and W{\"u}rzburg-Dresden Cluster of Excellence ct.qmat, Technische Universit{\"a}t Dresden, 01062 Dresden, Germany}
\author{Dominik~Kriegner}
\affiliation{Institut f{\"u}r Festk{\"o}rper- und Materialphysik and W{\"u}rzburg-Dresden Cluster of Excellence ct.qmat, Technische Universit{\"a}t Dresden, 01062 Dresden, Germany}
\affiliation{Institute of Physics, Czech Academy of Sciences, Cukrovarnick\'a 10, 162 00, Praha 6, Czech Republic}
\author{Philipp~Ritzinger}
\affiliation{Institut f{\"u}r Festk{\"o}rper- und Materialphysik and W{\"u}rzburg-Dresden Cluster of Excellence ct.qmat, Technische Universit{\"a}t Dresden, 01062 Dresden, Germany}
\author{Michaela Lammel}
\affiliation{Institute for Metallic Materials, Leibnitz Institute of Solid State and Materials Science, 01069 Dresden, Germany} 
\affiliation{Technische Universit\"{a}t Dresden, Institute of Applied Physics, 01062 Dresden, Germany}
\author{Miina Leiviskä}
\affiliation{Univ. Grenoble Alpes, CNRS, CEA, Grenoble INP, Spintec, F-38000 Grenoble, France}
\author{Vaclav~Pet\v{r}i\v{c}ek}
\affiliation{Institute of Physics, Czech Academy of Sciences, Cukrovarnick\'a 10, 162 00, Praha 6, Czech Republic}
\author{Petr Dole\v{z}al}
\affiliation{Department of Condensed Matter Physics, Faculty of Mathematics and Physics, Charles University, Ke Karlovu 5, 121 16 Prague 2, Czech Republic}
\author{Eva Schmoranzerova}
\affiliation{Department of Condensed Matter Physics, Faculty of Mathematics and Physics, Charles University, Ke Karlovu 5, 121 16 Prague 2, Czech Republic}
\author{Anton\'in Bad\hspace{-0.05cm}'\hspace{-0.05cm}ura}
\affiliation{Department of Condensed Matter Physics, Faculty of Mathematics and Physics, Charles University, Ke Karlovu 5, 121 16 Prague 2, Czech Republic}
\author{Andy Thomas}
\affiliation{Institut f{\"u}r Festk{\"o}rper- und Materialphysik and W{\"u}rzburg-Dresden Cluster of Excellence ct.qmat, Technische Universit{\"a}t Dresden, 01062 Dresden, Germany}
\affiliation{Leibniz Institute for Solid State and Materials Research Dresden (IFW Dresden), Institute for Metallic Materials, 01069 Dresden, Germany}
\author{Vincent~Baltz}
\affiliation{Univ. Grenoble Alpes, CNRS, CEA, Grenoble INP, Spintec, F-38000 Grenoble, France}
\author{Lisa~Michez}
\affiliation{Aix-Marseille Univ, CNRS, CINaM, Marseille, France}
\author{Jairo~Sinova}
\affiliation{Institut f\"ur Physik, Johannes Gutenberg Universit\"at Mainz, 55128 Mainz, Germany}
\affiliation{Institute of Physics, Czech Academy of Sciences, Cukrovarnick\'a 10, 162 00, Praha 6, Czech Republic}
\author{Sebastian~T.~B.~Goennenwein}
\affiliation{Institut f{\"u}r Festk{\"o}rper- und Materialphysik and W{\"u}rzburg-Dresden Cluster of Excellence ct.qmat, Technische Universit{\"a}t Dresden, 01062 Dresden, Germany}
\affiliation{Universit\"{a}t Konstanz, Fachbereich Physik,78457 Konstanz, Germany}
\author{Tom\'{a}\v{s}~Jungwirth}
\affiliation{Institute of Physics, Czech Academy of Sciences, Cukrovarnick\'a 10, 162 00, Praha 6, Czech Republic}
\affiliation{School of Physics and Astronomy, University of Nottingham, NG7 2RD, Nottingham, United Kingdom}
\author{Libor \v{S}mejkal}
\affiliation{Institut f\"ur Physik, Johannes Gutenberg Universit\"at Mainz, 55128 Mainz, Germany}
\affiliation{Institute of Physics, Czech Academy of Sciences, Cukrovarnick\'a 10, 162 00, Praha 6, Czech Republic}

\maketitle

{\bf Time-reversal ($\mathcal{T}$) symmetry breaking is a  fundamental physics concept underpinning a broad science and technology area, including topological magnets\cite{Smejkal2017b,Tokura2019,Smejkal2018}, axion physics\cite{nenno2020axion}, dissipationless Hall currents\cite{Nagaosa2010,Taguchi2001,Neubauer2009,Machida2010,Nakatsuji2015,Feng2020a}, or spintronic memories\cite{Chappert2007,MacDonald2011}. A best known conventional model  of macroscopic $\mathcal{T}$-symmetry breaking is a ferromagnetic order of itinerant Bloch electrons with an isotropic spin
interaction in momentum space \cite{POMERANCHUK1958,Herring1966,Landau9}. Anisotropic electron interactions\cite{POMERANCHUK1958,Wu2007}, on the other hand, have been a domain of correlated quantum phases, such as the $\mathcal{T}$-invariant nematics or unconventional  
superconductors \cite{Ii2007,Si2016}. Here we report discovery of a broken-$\mathcal{T}$ phase of itinerant Bloch electrons with an unconventional anisotropic spin-momentum interaction, whose staggered nature leads to the formation of two ferromagnetic-like valleys in the momentum space with opposite spin splittings. We describe qualitatively the effect by deriving a non-relativistic single-particle Hamiltonian model. Next, we identify the unconventional staggered spin-momentum interaction by  first-principles electronic structure calculations in a four-sublattice  antiferromagnet Mn$_{\text{5}}$Si$_{\text{3}}$ with a collinear  checkerboard magnetic order. We show that the staggered spin-momentum interaction is set by
nonrelativistic spin-symmetries\cite{Bradley} which were previously omitted in relativistic physics classifications of spin interactions and topological quasiparticles\cite{Ciccarelli2016,Cano2019}. Our measurements of a spontaneous Hall effect\cite{Nagaosa2010} in epilayers of antiferromagnetic Mn$_{\text{5}}$Si$_{\text{3}}$ with vanishing magnetization are consistent with our theory predictions. Bloch electrons with the unconventional staggered spin interaction, compatible with abundant low atomic-number materials, strong spin-coherence, and collinear antiferromagnetic order open unparalleled possibilities for realizing $\mathcal{T}$-symmetry broken spin and topological quantum phases.}

According to the Kramers theorem, an energy eigenstate of an electron and its time-reversed partner are degenerate in systems with $\mathcal{T}$-symmetry\cite{Kramers1930,Wigner1932}. On the microscopic level of individual atoms, breaking of the $\mathcal{T}$-symmetry by an applied magnetic field then leads to the Zeeman spin-splitting of the energy levels of the electron orbitals. The counterpart Zeeman effect in solids is the band spin-splitting around $\mathcal{T}$-invariant crystal momenta (TRIMs)\cite{Fu2007b}, where a momentum ${\bf k}$ is $\mathcal{T}$-invariant  when it differs from $-{\bf k}$ only by a reciprocal lattice vector.  Here the $\mathcal{T}$-symmetry breaking is macroscopic since, besides individual TRIMs,  the number of collectively  Zeeman spin-split electronic states is proportional to the macroscopic number of atoms forming the crystal. As a consequence, it can be detected by measuring macroscopic observables that are odd under $\mathcal{T}$. A textbook example among those is the ordinary Hall effect which is measured when a conductor is subjected to an applied magnetic field. 

In addition, the magnetic order of atoms in ferromagnets makes the macroscopic  $\mathcal{T}$-symmetry breaking  spontaneous. Here the Zeeman effect, illustrated in Fig.~1a,  occurs due to the internal electronic exchange coupling  rather than an externally applied magnetic field.  The macroscopic spontaneous $\mathcal{T}$-symmetry breaking in a conventional Stoner ferromagnet  is a  Fermi liquid instability in the isotropic ($s$-wave) spin channel  of interacting Bloch electrons\cite{POMERANCHUK1958,Herring1966,Landau9}. It can be probed in an electrical measurement by the anomalous Hall effect (AHE)  in the limit of zero magnetic field (also referred to as spontaneous Hall effect) \cite{Nagaosa2010}. 

Looking beyond ferromagnets,  the Zeeman effect at TRIMs is excluded in $\mathcal{T}$-symmetric non-magnetic systems. Similarly, the Zeeman-split bands, as well as  the spontaneous Hall effect and other macroscopic $\mathcal{T}$-symmetry breaking phenomena, have been also commonly considered to be excluded in antiferromagnets with an antiparallel alignment of atomic moments in the crystal\cite{Turov1965,Neel1971,Surgers2014,Surgers2016,Ghimire2018}.  The Coulomb-exchange field can be of comparable strength in antiferromagnets as in ferromagnets but is spatially localized. As a result, the $\mathcal{T}$-symmetry breaking spin-polarization can be visualized microscopically when focusing a probe on an individual atom\cite{Loth2012}. On the macroscopic level, however, the effect was assumed to cancel out due to the antiparallel alignment of the local Coulomb-exchange fields on neighboring atoms.  
Indeed, if the system has a symmetry in which  $\mathcal{T}$ is combined with a certain spatial symmetry operation connecting the opposite magnetic sublattices\cite{Smejkal2016}, the macroscopic $\mathcal{T}$-symmetry breaking is prohibited. An example is the class of antiferromagnets, whose $\mathcal{T}$ combined with spatial inversion ($\mathcal{P}$) 
results in Kramers degeneracy of electronic bands over the entire Brillouin zone. Recently, relativistic spin-orbit interaction\cite{Yamauchi2019} or anisotropic magnetization densities\cite{Smejkal2020,Ahn2019,Hayami2019,Yuan2020} were predicted to split bands in antiferromagnets, but only outside TRIMs in analogy with relativistic Rashba spin-splitting in non-centrosymmetric paramagnets\cite{Ishizaka2011}.

In this article, we demonstrate  a macroscopic $\mathcal{T}$-symmetry breaking in an  antiferromagnet with a  staggered Zeeman spin-splitting of TRIMs, as illustrated on a model with unconventional $d$-wave spin interaction in Figs.~1c-e. We start our description at the model level and then  provide an evidence of the anisotropic staggered Zeeman  interaction on the band structure of a collinear antiferromagnet Mn$_{\text{5}}$Si$_{\text{3}}$, calculated from first principles within the local density approximation. As a demonstration of the macroscopic $\mathcal{T}$-symmetry breaking by the unconventional staggered spin-momentum interaction, we calculate the intrinsic disorder-independent contribution to the AHE, proportional to the band structure Berry curvature\cite{Nagaosa2010}. Finally, we experimentally confirm the presence of AHE in epilayers of antiferromagnetic Mn$_{\text{5}}$Si$_{\text{3}}$  with a vanishing net magnetization and a remanent zero-field Hall conductivity on the scale of $\sim 10$~Scm$^{-1}$, consistent with theory.

{\bf\em Model Hamiltonian description.} In Fig.~1c-e we illustrate  the unconventional $d$-wave Zeeman effect on a model antiferromagnet with separated band structure valleys around two different TRIMs ${\rm\bf M}_{1}$ and ${\rm\bf M}_{2}$. The ${\rm\bf M}_{1}$ valley exhibits one sign of the Zeeman spin-splitting, and ${\rm\bf M}_{2}$  the opposite sign. The band structure is obtained by considering a 2D tight-binding two-band model in the second quantization basis (see Fig.~1d and Methods):
\begin{equation}
\mathcal{H}=\sum_{\langle i j\rangle} \widehat{c}_{i}^{\dagger}\left(t \sigma_{0}+ t_{J} {\bf d}_{ij} \cdot \boldsymbol\upsigma\right) \widehat{c}_{j}.
\end{equation}
The first term describes the  kinetic nearest-neighbor hopping parametrized by $t$ ($\sigma_0$ is a $2\times2$ identity matrix).  The second term is the unconventional $d$-wave Zeeman interaction due to the Coulomb-exchange dependent hopping, parametrized by $t_J$ ($\boldsymbol\upsigma$ are the spin Pauli matrices).  Physically, we can understand this term as an exchange molecular field felt by the electron when hopping on top of the collinear antiferromagnetic background. 
The sign of the exchange dependent hopping is positive (negative) along the $x$ ($y$) axis and is implemented by the unit vectors ${\bf d}_{ij}$ aligned with the N\'{e}el vector axis, as shown in Fig.~1d. We point out that this form of the hopping terms preserves the $\mathcal{P}$-symmetry of our model since hopping along positive and negative directions is the same. 

After performing the Fourier transformation into momentum space and the ${\bf k}\cdot{\bf p}$ approximation around TRIMs, we obtain an effective Hamiltonian (see Methods) for the ${\bf M}_{1}$ and ${\bf M}_{2}$ Zeeman spin-split valleys:
\begin{equation}
\mathcal{H}_{\text{eff}}({\bf k})=\epsilon({\bf k})+{\bf B}_{\text{AFZ,eff}}({\bf k})\cdot \boldsymbol\upsigma.
\label{HAF}
\end{equation}
Here the first term, $\epsilon({\bf k})=t(k_{x}^2-k_{y}^{2})\tau_{z}$,   is the kinetic energy of electrons with the crystal momentum {\bf k} measured from the respective ${\bf M}_{1(2)}$ TRIM and given in units of $1/a$ ($a$ is the lattice constant). 
The second term is the $d$-wave Zeeman interaction of the electron's spin with a field ${\bf B}_{\text{AFZ,eff}}=t_J\left(4-k^2\right)\tau_{z}\hat{{\bf e}}_{z}$. The pseudospin Pauli matrix $\tau_{z}$ describes the ${\bf M}_{1(2)}$ valley degree of freedom and $\hat{{\bf e}}_{z}$ is the unit N\'eel vector set along the $z$-axis. The $d$-wave Zeeman field, marked by magenta arrows in Fig.~1e, is opposite in the ${\bf M}_{1}$ and ${\bf M}_{2}$ valleys and couples only to the $z$-component of  spin. This makes both the valley ($\pm$) and spin ($\uparrow/ \downarrow$) indices good quantum numbers of the eigenstates of the model $d$-wave Zeeman  Hamiltonian (\ref{HAF}), as highlighted in Fig.~1e. The energy bands of the Hamiltonian (\ref{HAF}) calculated over the full Brillouin zone (see Methods) are shown in Fig.~1c. They illustrate the coexistence of the macroscopic spontaneous $\mathcal{T}$-symmetry breaking with the zero net moment and preserved $\mathcal{P}$-symmetry in our model antiferromagnet. The model has a $d_{x^2-y^2}$-wave symmetry of the spin interaction, corresponding to allowed collinear spin polarizations of Bloch states at TRIMs ${\bf M}_{1,2}$ with opposite spin splittings.  We emphasize that unlike the correlated phases \cite{Ii2007,Si2016},  the anisotropic $d$-wave interaction originates in our case purely from the Coulomb exchange. 

{\bf\em Ab initio calculations in Mn$_{\text{5}}$Si$_{\text{3}}$.} We now move on from the model Hamiltonian to the demonstration of the macroscopic $\mathcal{T}$-symmetry breaking  in the first-principles band structure of a  collinear antiferromagnetic phase of Mn$_{\text{5}}$Si$_{\text{3}}$ with the staggered  Zeeman spin-splitting of the TRIM valleys. The space group of a paramagnetic crystal of Mn$_{\text{5}}$Si$_{\text{3}}$ is $P6_{\text{3}}/mcm$ with a hexagonal unit cell containing two formula units.  In Fig.~2a we show that the sixteen atoms of the unit cell are occupying three different Wyckoff positions\cite{Gottschilch2012,Biniskos2018a}. These are 6g Mn$^{\text{(2)}}$ positions (black spheres), 4d Mn$^{\text{(1)}}$ sites (magenta spheres), and 6g Si positions (blue spheres). Neutron scattering data in bulk crystals showed that 4 out of the 6  Mn$^{\text{(2)}}$ sites in the unit cell are antiferromagnetically ordered in the temperature range of 70--100~K, while the remaining 6 Mn and 6 Si atoms are nonmagnetic\cite{Gottschilch2012,Biniskos2018a}. There are three different possible permutations of the 4 antiferromagnetic moments: Two stripy orderings\cite{Chaloupka2010} $ \begin{smallmatrix} \uparrow & \downarrow\\ \uparrow & \downarrow \end{smallmatrix}$ and$ \begin{smallmatrix} \uparrow & \uparrow\\ \downarrow & \downarrow \end{smallmatrix}$, and one checkerboard ordering $ \begin{smallmatrix} \uparrow & \downarrow\\ \downarrow & \uparrow \end{smallmatrix}$ with all nearest-neighbour interactions antiferromagnetic. On one hand, the first-two orderings of spins break the $\mathcal{T}$-symmetry but preserve a combined $\mathcal{PT}$-symmetry, resulting in Kramers degenerate bands with no spin-splitting allowed throughout the entire Brillouin zone. On the other hand, the $\mathcal{T}$-symmetry breaking checkerboard ordering, shown in Fig.~2a, breaks the $\mathcal{PT}$-symmetry. This opens the possibility for generating the macroscopic $\mathcal{T}$-symmetry breaking  in the band structure, while preserving  a zero net magnetization. Note that a triangular non-collinear antiferromagnetic ordering of the Mn$^{\text{(2)}}$ atoms is unfavorable in Mn$_{\text{5}}$Si$_{\text{3}}$ due to the different lengths of the Mn$^{\text{(2)}}$ bonds\cite{Brownt1992}. (This also applies to a thin film geometry relevant for our experiment.)

Our density-functional theory  (DFT) calculations in the local density approximation show that for the collinear antiferromagnetic checkerboard phase of Mn$_{\text{5}}$Si$_{\text{3}}$, the states around the Fermi level comprise mainly of perfectly compensated itinerant antiferromagnetic Mn $d$ states (for details see Methods and Supplementary Fig.~S2). The antiferromagnetic state is stable in our local-density-approximation calculations even without introducing the Hubbard corrections and has lower total energy than the ferromagnetic and paramagnetic states by 0.22 and 0.71~eV per unit cell, respectively.  Fig.~2a shows magnetization density isosurfaces calculated without relativistic spin-orbit coupling, which are nearly isotropic around the atomic sites. 

In the inset of Fig.~2b, we show the hexagonal Brillouin zone with the notation of high symmetry points, out of which eight are TRIMs\cite{Fu2007b}: $\boldsymbol\Gamma$, ${\bf M}$, ${\bf M}_{1}$, and ${\bf M}_{2}$ in the $k_{z}=0$ plane and $
{\bf A}$, ${\bf L}$, ${\bf L}_{1}$, and ${\bf L}_{2}$ in the $k_{z}=\pi$ plane.  Our non-relativistic DFT calculations of spin projected energy bands are shown in Fig.~2b. The $\boldsymbol\Gamma$, ${\bf A}$, ${\bf M}$, and ${\bf L}$ points are spin-degenerate as one would commonly expect for a collinear antiferromagnetic ordering. In stark contrast, and consistent with our minimal model in Eq.~\eqref{HAF}, we observe staggered Zeeman splittings at TRIMs ${\bf M}_{1}$, $-{\bf M}_{2}$, ${\bf L}_{1}$, and $-{\bf L}_{2}$. Here TRIMs  ${\bf M}_{i}$ and $-{\bf M}_{i}$ (and similarly ${\bf L}_{i}$ $-{\bf L}_{i}$) are equivalent for both  $i=1,2$ due to the presence of the $\mathcal{P}$-symmetry in the crystal. 

The  Mn$_{\text{5}}$Si$_{\text{3}}$ antiferromagnet has many metallic bands where different Zeeman splittings overlap. In Fig. 2b we illustrate the staggered Zeeman splitting in the valence bands along  $\boldsymbol\Gamma$${\bf M}_{1}$${\bf K}$$-{\bf M}_{2}$$\boldsymbol\Gamma$ and ${\bf A}$${\bf L}_{1}$${\bf H}$$-{\bf L}_{2}$${\bf A}$ lines by grey shading. We marked the strongest Zeeman spin splitting in the chosen energy window. The Zeeman fields with the alternating sign at the two non-equivalent valleys are highlighted by purple arrows.
(We plot a larger energy window in Supplementary Fig.~S2.)  

The anisotropic staggered nature of the spin-momentum interaction is visible in Fig.~2c on the $k_{z}=0$ cut of the Fermi surface. It is linked to non-relativistic symmetry operations in the antiferromagnet in which spin-space rotations are decoupled from symmetry operations  in the real space\cite{Bradley}. The construction of spin symmetry groups of collinear antiferromagnets is defined and discussed in detail in the Supplementary information.  In our collinear antiferromagnetic phase of Mn$_{\text{5}}$Si$_{\text{3}}$, the spin symmetry point group  contains a real-space $\mathcal{P}$-symmetry, and combined spin-reversal and real-space mirror symmetries  $\mathcal{R}_{S}\mathcal{M}_{x}$ and $\mathcal{R}_{S}\mathcal{M}_{y}$, with the crystal mirror planes $\mathcal{M}_{x(y)}$ shown in Fig.~2a. (Note that these symmetries were already present in our model of the unconventional spin $d$-wave interaction, as shown in Fig.~1.) The $\mathcal{R}_{S}\mathcal{M}_{x(y)}$  symmetry is also contained in the little spin symmetry group of the $k_{x(y)}=0$ surface, imposing spin degeneracy at these wavevectors, i.e., also at the $\boldsymbol\Gamma$ TRIM.  On the other hand, the little spin symmetry group of the ${\bf M}_{1}$ and ${\bf M}_{2}$ wavevectors contains no symmetry operation involving $\mathcal{R}_{S}$, which makes the spin splitting allowed by symmetry at these TRIMs. Finally, the $\mathcal{P}$ and $\mathcal{R}_{S}\mathcal{M}_{x(y)}$ symmetries of our collinear antiferromagnetic phase of Mn$_{\text{5}}$Si$_{\text{3}}$ ensure that the staggered spin-momentum interaction generates oppositely spin-split valleys  around ${\bf M}_{1}$ and ${\bf M}_{2}$ TRIMs which perfectly compensate each other (see Supplementary information for explicit proofs). 

In  Supplementary Fig.~S2 and S3, we plot the corresponding Fermi surface cut and energy bands calculated with relativistic spin-orbit coupling. We set the N\'eel vector along the [111]-direction (in Cartesian coordinates) which corresponds to lower total energy than a N\'eel vector in the $x-y$ plane or along the $z$-axis (see Methods and Supplementary Fig.~S4 for the magnetic anisotropy calculations). We observe that the effect of the spin-orbit coupling on energy scales of the valleys and their Zeeman splitting is negligible, owing to the combined effect of light Mn and Si elements and the Coulomb-exchange origin of the staggered spin-momentum interaction. The spin-orbit coupling lowers the number of symmetries but only weakly breaks the perfect compensation among the valleys with opposite Zeeman spin splitting, as illustrated in Supplementary Figs.~S2 and S3. 

We also observe that the Zeeman split bands in the valleys remain nearly perfectly spin-polarized. A weak spin rotation away from the N\'eel vector axis occurs only in small regions of spin-orbit hybridized bands, marked by lower intensity color in Supplementary Figs.~S3.  The comparison of bands calculated with and without spin-orbit coupling  illustrates that analyzing the relativistic magnetic symmetry groups cannot distinguish the microscopic spin-orbit and Coulomb-exchange mechanisms of splittings. This might also explain why the macroscopic spontaneous $\mathcal{T}$-symmetry breaking in a staggered  Zeeman spin-split antiferromagnet  has been missed in the literature to date\cite{Ciccarelli2016,Cano2019}. The exhaustive classifications of spin-orbit fields\cite{Ciccarelli2016} and magnetic symmetry groups\cite{Cano2019} prevent the decoding of this unconventional symmetry breaking mechanism. We also note that the spin symmetries in the anisotropic staggered Zeeman-split  antiferromagnet Mn$_{\text{5}}$Si$_{\text{3}}$ lead to additional unique features in the electronic structure such as a degeneracy of the bands with the same spin and a linear dispersion at ${\bf L}_{1}$ and ${\bf L}_{2}$ points, shown in Supplementary Fig.~S2.

{\bf\em Topological Berry curvature and anomalous Hall effect.} We now demonstrate that the macroscopic $\mathcal{T}$-symmetry breaking arising from the unconventional staggered spin-momentum interaction represents a new route to the generation of topologically nontrivial quantum states. The topological character of the Bloch states manifests itself in a nonzero Berry curvature field of a vector bundle of Bloch states $u(\textbf{k})$ in the momentum space, 
$
\boldsymbol\Omega(\textbf{k})=-\text{Im} \langle\partial_{\textbf{k}}u(\textbf{k}) \vert \times \vert \partial_{\textbf{k}}u(\textbf{k}) \rangle
$.
In Fig.~2d, we plot our Berry curvature calculations for a Wannier model of antiferromagnetic Mn$_{\text{5}}$Si$_{\text{3}}$, derived from the relativistic DFT theory (see Methods). Here, the relativistic spin-orbit coupling connects the spin and momentum spaces, in analogy to Berry curvature and AHE studies in ferromagnets\cite{Nagaosa2010,Gosalbez-Martinez2015}. For the N\'eel vector along the [111] direction, all three components $\Omega_{x}$, $\Omega_{y}$, and $\Omega_{z}$ are nonzero and exhibit sizable hotspots. The Berry vector is consistent with the low magnetic symmetry group $\overline{1}$ (containing only identity and inversion) of Mn$_{\text{5}}$Si$_{\text{3}}$ with spin-orbit coupling and the [111]-oriented N\'eel vector. In Fig.~2e, we show the Berry curvature integrated over the Brillouin zone which is proportional to the intrinsic contribution (independent of scattering) of the anomalous Hall conductivity  $\sigma_{xy}$\cite{Gosalbez-Martinez2015,Nagaosa2010}. Our calculations  illustrate that the Hall conductivity can reach values comparable to typical ferromagnets\cite{Nagaosa2010}, and we predict a sizable $\sigma_{xy}\sim5-20$~Scm$^{-1}$ within a $\sim 100$~meV energy window around the Fermi level (see also Supplementary Fig.~S2).

In the remaining part of the paper, we present our experimental structural, magnetization and magneto-transport study of epitaxial thin-films of Mn$_{\text{5}}$Si$_{\text{3}}$. The measurements confirm the macroscopic spontaneous $\mathcal{T}$-symmetry breaking in this compensated antiferromagnet, consistent with our DFT calculations. We have grown 12~nm thick epitaxial layers of Mn$_{\text{5}}$Si$_{\text{3}}$(0001) by molecular beam epitaxy on top of a Si(111) substrate (see Methods for the growth description and Supplementary Information for detailed sample characterization). 
In Fig.~3a, we present a transmission electron microscopy image showing the orientation of our Mn$_{\text{5}}$Si$_{\text{3}}$ films on the Si substrate (see also Methods and Supplementary Fig.~S5 and S6). The inset of Fig.~3a shows the corresponding side view of the Mn$_{\text{5}}$Si$_{\text{3}}$ unit cell. Our X-ray diffraction  (XRD) scans at room temperature are consistent with an in-plane hexagonal symmetry of the epilayers (see Supplementary Information for more details). 

The longitudinal resistivity shown in Fig.~3b indicates a metallic character and its magnitude is similar to the one reported in thicker sputtered layers of Mn$_{\text{5}}$Si$_{\text{3}}$\cite{Surgers2016}. Previous reports on bulk and thicker layers identified \cite{Lander1967,Brownt1992,Brown1995,Gottschilch2012,Biniskos2018a,Surgers2014,Surgers2016} a non-coplanar antiferromagnetic phase AF$_1$ below a transition temperature $T_{\text{N1}} \approx 70$~K and a collinear antiferromagnetic phase AF$_2$ between $T_{\text{N1}}\approx70$~K and  $T_{\text{N2}}\approx100$~K. In our samples, we indeed observe two kinks in the longitudinal resistivity, highlighted by plotting the temperature derivative of the resistivity in Fig.~3b, signalling the transitions between the different phases.  
The first transition temperature corresponds to $T_{\text{N1}}$ quoted above. Remarkably, we observe a significant enhancement of $T_{\text{N2}} \approx 240$~K in our epitaxial thin-films.
We attribute the enhancement of $T_{\text{N2}}$ to a sizable strain in the epilayers as evidenced by our temperature dependent XRD and TEM studies (see also Supplementary Fig.~S6). 
The temperature dependence of the lattice constants obtained from the XRD measurements is shown in Fig.~3c. The lattice constant $a_{\rm Si}$ of the cubic silicon substrate exhibits only a weak temperature dependence (grey line in Fig.~3c). 
The out-of-plane lattice constant $c$ of the Mn$_{\text{5}}$Si$_{\text{3}}$ epilayer is in the whole temperature range smaller than the corresponding values obtained in bulk crystals \cite{Gottschilch2012}. 
This demonstrates the presence of epitaxial strain which can stabilize the antiferromagnetic ordering in our epilayers up to higher temperatures\cite{Ney2005}. In Fig.~3c, we see that the collinear antiferromagnetic phase in bulk Mn$_{\text{5}}$Si$_{\text{3}}$ is stabilized for $c$ smaller than  $\approx4.8$~\AA\,  (dotted line in Fig.~3c) which corresponds to a temperature range between 70 and 100~K, marked as AF$_{2,{\rm bulk}}$ in Fig.~3c. Our strained epilayers exhibit a lattice constant  $c$ smaller than $\approx4.8$~\AA\,  in a wider temperature window marked as AF$_{2}$ in Fig.~3c. This potentially explains the enhanced $T_{\text{N2}}$.

In the temperature range of 80 to 300~K, the measured out-of-plane lattice parameter $c$ of our films exhibits only a weak change of its linear slope near $T_{\text{N2}}$ without abrupt changes (see Fig.~3c). This indicates that the hexagonal unit cell is preserved over this wide temperature range and suggests that our epilayers do not exhibit the antiferromagnetic doubling of the unit cell seen in bulk samples below $T_{\text{N2}}$\cite{Brown1995,Gottschilch2012} and strain-relaxed single crystals\cite{Surgers2017}. The antiferromagnetic doubling makes the system invariant under a combined $\boldsymbol{t}\mathcal{T}$-symmetry ( $\boldsymbol{t}$ marks partial unit cell translation) which explains the reported absence of the AHE between $T_{\text{N1}}$ and $T_{\text{N2}}$ in bulk crystals\cite{Surgers2017}.  The unit cell doubling in bulk is accompanied by an orthorhombic symmetry lowering which, as mentioned above, is not observed in our epilayers. Correspondingly, we do not expect the {\bf t}$\mathcal{T}$-symmetry in our epilayers in the antiferromagnetic phase below $T_{\text{N2}}$ which opens the possibility for observing the AHE signature of macroscopic $\mathcal{T}$-symmetry breaking in our  staggered Zeeman-split antiferromagnet. Below, we experimentally confirm this expectation and show that the measured anomalous Hall signal and vanishing magnetization are consistent with the collinear checkerboard ordering considered in the DFT calculations. 

To perform the magnetotransport measurements, we have lithographically patterned the epilayers into Hall bars as shown in Fig.~3d (see Methods for more details on device fabrication). We simultaneously measure the longitudinal magnetoresistance with ordinary field dependence shown in Fig.~3e and total Hall resistivity $\rho_{xy}^{\text{tot}}$ in an external magnetic field $B_{z}$ applied along the surface normal, parallel to [0001] (See Supplementary Fig.~S7 for raw data). By fitting the linear-in-field ordinary Hall resistivity contribution, we determine the ordinary Hall coefficient, $R_{\rm H}\approx1-4\times10^{-10}$~m$^{3}$C$^{-1}$, which corresponds to a metallic carrier density, $n\sim10^{22}$~cm$^{-1}$ assuming a single-band model.
In Fig.~4a, we show the Hall resistivities after subtracting the linear ordinary Hall effect signal as $\rho_{xy}^{AF}=\rho_{xy}^{\rm tot}-R_{\rm H}B_{z}$ (see Methods and Supplementary Fig.~S7 and S8). We observe a sizable anomalous Hall resistivity below $T_{\text{N2}}\sim 240$~K, confirming the presence of  magnetic ordering. The anomalous Hall resistivity $\rho_{xy}^{AF}(B_{z})$ exhibits a large coercivity of $\approx 2$~T, which is also consistent with the presence of the antiferromagnetic order in our weakly spin-orbit coupled Mn$_{\text{5}}$Si$_{\text{3}}$. 
Moreover, our magnetometry measurements show a negligible remanent magnetization of a magnitude below  the detection limit of $\sim 0.01$~$\mu_{B}$  per unit cell, as seen in Fig.~4b, again consistent with the picture of a compensated antiferromagnetic phase below
$T_{\text{N2}}$.

In Fig.~4c, we decompose the antiferromagnetic Hall resistivity in two contributions:
$
\rho_{xy}^{AF}=\rho_{xy}^{AFZ}+\rho_{xy}^{THE}.
$
The $\rho_{xy}^{THE}$ component appearing below $T_{\rm N1}\approx70$~K with a bump-like hysteresis and reaching 0.09~$\mu\Omega$cm is consistent with the topological Hall effect due to a non-coplanar spin structure\cite{Neubauer2009,Surgers2014,Surgers2016,Surgers2017}, as reported previously for sputtered layers of Mn$_{\text{5}}$Si$_{\text{3}}$ with $\rho_{xy}^{THE}\approx0.02-0.04$~$\mu\Omega$cm \cite{Surgers2014}. The transition from a collinear antiferromagnetic phase above $T_{\rm N1}$ to an antiferromagnetic order with an additional non-coplanar spin component below $T_{\rm N1}$  is corroborated\cite{Gopalakrishnan,Taylor2020,Higo2018} in our films also by the observed enhancement of the longitudinal magnetoresistance (Fig.~3e). 
In contrast, the high temperature magnetic phase exhibits very small magnetoresistance further corroborating the presence of collinear antiferromagnetism. We remark that the  magnetoresistance in the low temperature phase is sizable even for smaller applied fields $\sim$ 0.5~T where the topological Hall effect is maximal. The $\rho_{xy}^{AFZ}$  contribution to the measured Hall resistivity observed below $T_{\rm N2}\approx240$~K is ascribed to the collinear antiferromagnetic projection of the moments even in the low temperature non-coplanar phase, and dominates the signal over the entire temperature range of $10-240$~K (see Figs.~4c,d).
The spontaneous value at remanence reaches 0.2--0.7~$\mu\Omega$cm (see also Supplementary Fig.~S9 for data measured on other samples).  

In Fig.~4e, we show the anomalous Hall conductivity $\sigma_{xy}^{AFZ}\approx\rho_{xy}^{AFZ}/\rho_{xx}^{2}(B_{z}=0)$ in the whole temperature range in which the samples are magnetically ordered. The $\sigma_{xy}^{AFZ}$ magnitude reaches values between 5 and 20 Scm$^{-1}$ (see also Supplementary Fig.~S9). This is
consistent with our DFT calculations of the anomalous Hall conductivity of Mn$_{\text{5}}$Si$_{\text{3}}$ with the checkerboard magnetic ordering. 
We remark that in polycrystalline films the Hall resistivity can also vanish due to the compensation from domains with the opposite N\'{e}el vector unlike in our epilayers\cite{Surgers2014,Surgers2016}. In Supplementary Fig.~S9 we show that also in our films the Hall resistivity magnitude decays with lowering the crystal quality.

{\bf\em Discussion.} Macroscopic $\mathcal{T}$-symmetry breaking arising from the unconventional staggered spin-momentum interaction opens an uncharted territory in spin physics and applications. In contrast to the correlated phases, the anisotropic staggered Zeeman effect  occurs on the lowest exchange-approximation level of electron-electron interactions and, as illustrated on our itinerant antiferromagnet Mn$_{\text{5}}$Si$_{\text{3}}$, can be identified using standard DFT methods with the local density approximation. This makes the anisotropic staggered Zeeman effect principally as robust as conventional $s$-wave ferromagnetism, and will facilitate an extensive computational research of  suitable material candidates within the broad family of antiferromagnets.

Among the new perspectives  in spin physics opened by the unconventional staggered Zeeman effect, we point out that TRIM states in centrosymmetric magnets are known to encode the information on whether the system can realise topological magnetic phases of matter\cite{Turner2012}. The  staggered Zeeman effect at TRIMs is, therefore, directly relevant for the realization of axion insulators, topological magneto-electrics, Weyl fermions or the quantum AHE\cite{Smejkal2017b}.

From a more applied perspective, we point out that the anisotropic staggered spin-momentum interaction opens new possibilities for efficient spin-charge conversion in spintronics information technologies. For example, it can facilitate the realization of the long-sought antiferromagnetic analogues\cite{MacDonald2011} of giant-magnetoresistance and spin transfer torque phenomena  which underpin reading and writing information in commercial magnetic random access memories \cite{Chappert2007}.

Besides ferromagnetism, the second branch of conventional spin-splitting physics has been based on the relativistic spin-orbit coupling. Compared to this area, we point out that the magnitude of the staggered Zeeman spin-splitting in Mn$_{\text{5}}$Si$_{\text{3}}$ is as large as 0.2 eV. This is a value corresponding to the record values reported for relativistic Rashba splittings in bulk materials with heavy elements, such as $\sim$0.1 eV in BiTeI 
(see Supplementary Information Tab.~1)\cite{Ishizaka2011}. Moreover, unlike the relativistic spin-orbit splitting, the  collinear exchange mechanism of the anisotropic staggered Zeeman effect conserves spin and does not require complex locally or globally non-centrosymmetric crystals and more exotic (and often toxic) heavy elements. Its intrinsic exchange origin makes it also stronger than the electric-gating induced splittings in the antiferromagnet MnPSe$_{\text{3}}$ \cite{Sivadas2016}, or the splittings by external magnetic fields  \cite{Ramazashvili2009}. From an opposite perspective, the high coercive fields observed in our experiments on Mn$_{\text{5}}$Si$_{\text{3}}$ also illustrate the protection of electron transport phenomena associated with our macroscopic spontaneous $\mathcal{T}$-symmetry breaking against external magnetic fields. In comparison, e.g., enhanced robustness of Cooper pairs against magnetic fields was induced externally by electrical gating in a superconducting transistor structure\cite{Lu2015a}.

\newpage
\section{Methods}

\textbf{Antiferromagnetic Zeeman effect tight-binding model.}
The full Hamiltonian of the square lattice antiferromagnetic Zeeman effect in crystal momentum space reads:
\begin{equation}
\mathcal{H}(\textbf{k})=2t\left( \cos {k_{x}}+ \cos {k_{y}} \right)\boldsymbol{1} + 2t_{J}\left( \cos {k_{x}}- \cos {k_{y}} \right) \boldsymbol\sigma\cdot \textbf{d}.
\end{equation}
Here $\textbf{d}$ is a unit vector along the N\'{e}el vector, and $\boldsymbol{1}$ is the unit matrix. 
The energy bands of the perfectly spin up and down polarized band are obtained as:
\begin{equation}
E_{\uparrow/\downarrow}(\textbf{k})=2t\left( \cos {k_{x}}+ \cos {k_{y}} \right)\pm 2t_{J}\left( \cos {k_{x}}- \cos {k_{y}} \right).
\end{equation}
We show the full 3D band structure of the model in Fig.~1c. Remarkably, the band connectivity of our model (as also highlighted in Supplementary Fig.~S1a) corresponds to the shaded bands of Mn$_{\text{5}}$Si$_{\text{3}}$ in Fig.~2b.  

\textbf{Magnetic and spin symmetry group analysis.}
The spin symmetries discussed in the main text enforce spin degenerate bands marked by dotted lines in the Fermi surface shown in Fig.~S3. We discuss the detailed formulation of spin symmetries in the Section III of Supplementary information. 
The magnetic space groups of the collinear antiferromagnetic hexagonal unit cell used in the main text are of type-III\cite{Bradley}. A N{\'e}el vector along [001] corresponds to $Cmcm$ 
(magnetic point group $mmm$) which enforces the Hall conductivity to vanish. The N{\'e}el vector oriented along [111] gives $P\overline{1}$ (magnetic point group $\overline{1}$) and all three components of Hall conductivity allowed as explained in the main text. Finally, an in-plane N{\'e}el vector parallel to the [100] direction crystallizes in $Cmc'm'$  
(magnetic point group $m^{\prime}m^{\prime}m$ with antiunitary rotation around out-of-plane-axis $\mathcal{C}_{2z}\mathcal{T}$) which allow for Hall conductivity components $(\sigma_{yz},\sigma_{zx},0)$. The magnetic space group symmetry of the bulk collinear antiferromagnetic phase of Mn$_{\text{5}}$Si$_{\text{3}}$ is $P_{c}bcm$ (magnetic point group $mmm1^{\prime}$) \cite{Gottschilch2012}. The symmetry group is of black-and-white type-IV\cite{Bradley} with an antiferromagnetic doubling vector along the $(\frac{1}{2}, \frac{1}{2}, 0)$ direction which excludes the anomalous Hall conductivity in bulk\cite{Smejkal2020}. From the above, only one phase is consistent with the experimentally observed Hall conductivity $\sigma_{xy}$, namely the collinear antiferromagnetic hexagonal phase with the N{\'e}el vector along [111] Cartesian direction.

\textbf{Antiferromagnetic density functional theory calculations.}
The density functional theory calculations were performed using the VASP package \cite{Kresse1996a} employing the projector augmented plane wave method \cite{Blochl:1994aa}. 
We have set the energy cut-off of the plane wave basis at 520 eV, used the PBE exchange-correlation functional \cite{Perdew1997}, and the wavevector grid 9 $\times$ 9 $\times$ 12. We have used the in-plane high-temperature lattice constant a=6.902A\cite{Gottschilch2012} and the c-lattice constant corresponding to the bulk collinear phase at T=70 K and our epilayers at T=170 K (4.795A). 
Our ground state energy calculations confirm the preference of a low symmetry orientation of the N{\'e}el vector over in-plane orientation and out-of-plane orientation by 0.9 and 0.2 meV/unit cell, respectively (See Supplementary Fig.~S4). We have calculated the out-of-plane net moment 0 and  0.03~$\mu_{B}$ per unit cell for the N{\'e}el vector along [001] and [111], respectively. In summary, our calculations indicate the orientation of the N{\'e}el vector slightly tilted out-of-the [001] axis.

\textbf{Berry curvature calculations of Hall conductivity.}
We have constructed a maximally localized Wannier function and the effective tight-binding model by using the Wannier90 code \cite{Mostofi2008}.
We have calculated the intrinsic anomalous Hall conductivity in WannierTools Python package\cite{Wu2017b} by employing the Berry curvature formula. We have used a fine-mesh of 320 $\times$ 320 $\times$ 240 Brillouin zone sampling points and have checked the convergence. The DFT calculated Hall conductivity confirms the symmetry analysis, e.g. all three components nonzero for the N{\'e}el  vector oriented along [111]. The experimentally observed component is $\sigma_{xy}$. The other two components can take even larger values of $\sim$400 S/cm (in dependence on energy) as we show in Fig.S4.  

\textbf{Epitaxial crystal growth.}
We have grown the epilayers by ultrahigh-vacuum molecular beam epitaxy (MBE) with a base pressure less than $10^{-10}$ Torr. 
We have cleaned the Si(111) substrate surface by using a modified Shiraki method \cite{ishizaka1986low}. We have formed a final oxide layer chemically to protect the Si surface against oxidization in ambient air. This thin oxide layer was then thermally removed by annealing at 900$^{\circ}$C during a few min in the MBE chamber. Subsequently, a 10~nm-thick Si buffer layer was deposited at 600$^{\circ}$C to ensure a high-quality starting surface. The surface of the sample was monitored in situ by the reflection high energy electron diffraction (RHEED) technique that revealed an atomically flat surface with a well-developed (7$\times$7) reconstruction (see Supplementary Fig.~S5) We decreased the growth temperature to $170^{\circ}C$ for the subsequent deposition of Mn and Si. We have evaporated high-purity Mn and Si by using a conventional high-temperature effusion sublimation cells. We have calibrated the cell fluxes by using RHEED oscillations and a quartz microbalance to achieve the desired stoichiometry of the layers with a total growth rate in the range of 0.1 - 0.2 \AA/s.  The first monolayers exhibited the typical signature of a Mn$_5$Si$_3$-type crystal, a $(\sqrt{3} \times \sqrt{3})$R$30^{\circ}$ reconstruction\cite{OLIVEMENDEZ2008191}. Crystal quality was further improved by thermal annealing with its quality degree monitored by RHEED pattern (see Supplementary Fig.~S5). Different growth parameters (including the nominal thickness the Mn/Si layers, the Mn and Si deposition rate and the growth temperatures) were optimized to minimize the presence of the spurious MnSi phase. We note that the Curie temperature of MnSi is around 30K and therefore, cannot contribute to the measured signal up to 240K. The same is valid for typical Mn-based oxides which have typically low critical temperature. We show the amount of the spurious phase in our five different samples in Supplementary Fig.~S9. 

\textbf{Transmission electron microscopy and X-ray diffraction.}
TEM investigations were performed at an accelerating voltage of 300 kV on a JEOL JEM-3010 instrument with a spatial resolution of 1.7 Å. The transmission electron microscopy (TEM) cross-section specimens were prepared by using a dual focused ion beam (FEI Helios 600 NanoLab) milling through a liftout technique.
The TEM analyses summarized in Supplementary Fig.~S5 confirmed the epitaxial relationships (Mn$_{\text{5}}$Si$_{\text{3}}$(0001)[10-10]//MnSi(111)[1-10]//Si(111)[11-2]) and reveals the location of MnSi at the interface between the Si substrate and Mn$_5$Si$_3$. The lattice mismatch of 3.7 percent between Si(111) and Mn$_5$Si$_3$ is accommodated by the formation of a thin layer of interfacial MnSi and an array of interfacial dislocations. XRD measurements at room temperature were realized using a high brilliancy rotating anode, Rigaku RU-200BH equipped with an image plate detector, Mar345. The radiation used was Cu K$\alpha$, $\lambda$ = 1.5418Å and the beam size was $0.5 \times  0.5$ mm$^2$. 
The high-intensity Mn$_{\text{5}}$Si$_{\text{3}}$ 0002 reflection in the XRD data recorded at 300K and shown in Supplementary Fig.~S6 evidenced the preponderant formation of the Mn$_5$Si$_3$  hexagonal phase that grows along the c-axis. 
Low-temperature XRD experiments were performed in the Bragg-Brentano 
geometry using a Siemens D500 diffractometer. Cooling of the sample was provided by a closed-cycle refrigerator (CCR, Sumitomo Heavy Industries), and He exchange gas ensured equalization of the temperature between the cold-finger, thermometer and sample. Cu-K$_{\alpha1,2}$ radiation and a linear detector were used to speed up the data recording \cite{Kriegner:po5035}.

\textbf{Magnetotransport and anomalous and topological Hall extraction.}
We have patterned the Hall bars by standard optical lithography and Argon plasma etching.  
In Supplementary Fig.~S7, we show the raw transversal and longitudinal resistivity data, measured simultaneously. 
In Supplementary Fig.~S8, we show the subtraction of the linear slope, i.e. the 
ordinary Hall effect. The measured data were separated into symmetric and antisymmetric component. In Fig.4a we show only the antisymmetric part.
This procedure removes the small constant offset in transverse resistivity caused by tiny misalignments of the Hall contacts and the even contribution to the transverse signal. The even contribution presumably originates from anisotropic magnetoresistance due to the low symmetry\cite{Seemann2015}. 
The anomalous and topological Hall resistivities were extracted by fitting a cosh function. The anomalous Hall contribution is taken as the amplitude of the cosh fit. We show the result in Fig.~S8. The additional bump-like features correspond to the topological Hall signal \cite{Surgers2014}. The recalculated amplitude of the AHE reaches 5-20 S/cm and correlates with the quality of the crystal, as shown in Supplementary Fig.~S9. 

\textbf{Magnetometry measurements.}
For the magnetic characterization of the Mn$_5$Si$_3$ thin films, a Quantum Design MPMS7-XL SQUID magnetometer with reciprocrating sample option has been used. The unpatterned sample was cleaned prior to the measurement and mounted using plastic straws. The field dependent magnetization has been measured at different temperatures for magnetic field strengths between $\pm$5 T (cp. Fig.~4b) applied out of the sample plane. The signal is dominated by the diamagnetism of the silicon substrate, this diamagnetic contribution is, however, negligible in the small magnetic field (inset). %

\section{Acknowledgement}
RS andd STBG acknowledge financial support by the Deutsche
Forschungsgemeinschaft (DFG) via SFB 1143/C08 and GO 944/8-1. The low temperature X-ray diffraction was performed in MGML (http://mgml.eu/), which was supported within the program of Czech Research Infrastructures (project no. LM2018096). ES acknowledges INTER-COST grant no. LTC20026. This work was supported by the French national research agency (ANR) (Project ASTRONICS - Grant Number ANR-15-CE24-0015-01; and Project MATHEEIAS), and the CNRS International Research Project (IRP) program (Project SPINMAT). V.P.  acknowledges the Czech Science Foundation project no. 18-10504S. T.J. acknowledges Ministry of Education of the Czech Republic Grants LNSM-LNSpin, LM2018140,
Czech Science Foundation Grant No. 19-28375X, and the Neuron Endowment Fund Grant. TJ, JS and LS acknowledges the EU FET Open RIA Grant No. 766566. JS and LS acknowledge SPIN+X (DFG SFB TRR 173) and Elasto-Q-Mat (DFG SFB TRR 288). R.G.H. and L.S. gratefully acknowledge the computing time granted on the supercomputer Mogon at Johannes Gutenberg University Mainz (hpc.uni-mainz.de).

\section{Author contributions}
The project was conceived and led by H.R. and L.S.
 The sample growth and structural characterization and analysis was performed by L.M., I.K., D.K., V.P., R.L.S., R.S., H. R. fabricated devices D.K., P.D. performed the structural characterization at low temperature R.L.S., H.R., M.L. performed the magnetometry measurements. H.R., R.L.S., R.S., A.B., P.R., E.S. performed the magneto-transport measurements at various samples in various setups. H.R., R.L.S., S.T.G.B., D.K., V.B., A.T. analyzed the magneto-transport measurements. The density functional theory calculations and theoretical analysis was performed by R.G.H, J.S., T.J. and L.S. The model spin d-wave antiferromagnet was developed by L.S. and T.J. The manuscript was written by L.S., H.R. and T.J. All authors discussed the experimental and theoretical data and commented on the manuscript. 

\bibliographystyle{naturemag}

\newpage

\begin{figure}[h]
\hspace*{0cm}\epsfig{width=1\columnwidth,angle=0,file=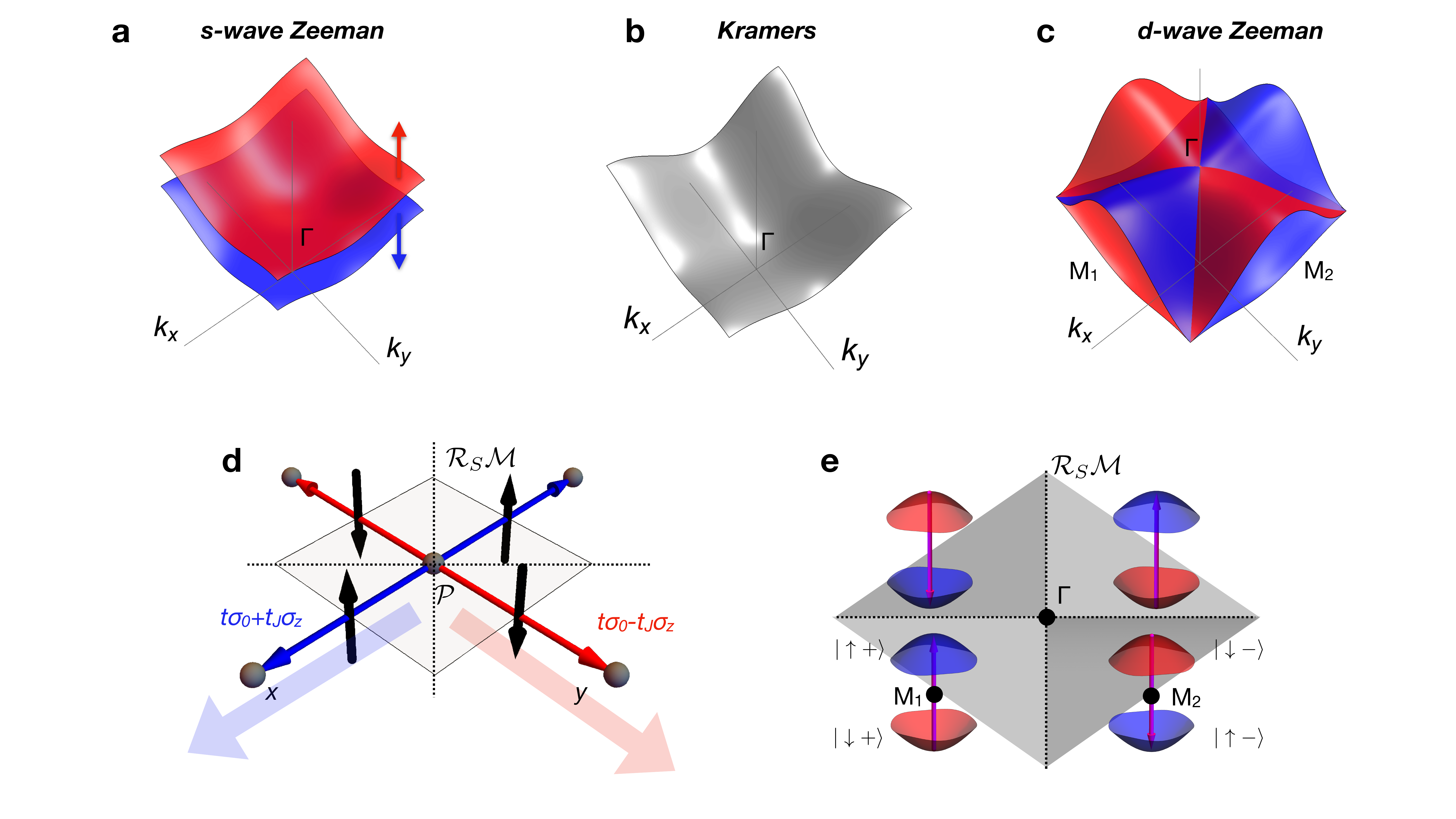}
  \caption{\textbf{Macroscopic $\mathcal{T}$-symmetry breaking arising from a model spin d-wave interaction.}
 \textbf{a} Zeeman spin-split bands at the $\mathcal{T}$-invariant momentum $\Gamma$ of a model ferromagnet with spin s-wave interaction. The spin-up and spin-down bands are marked by the red and blue colour, respectively. 
 \textbf{b} Kramers degenerate bands of a $\mathcal{PT}$-symmetric antiferromagnets and non-magnets. 
  \textbf{c} Staggered spin d-wave interaction in an antiferromagnet: Blue and red colored  bands  correspond to a perfect spin-up and spin-down polarization, respectively.   
  \textbf{d} Schematics of a minimal tight-binding model of the spin d-wave interaction. The kinetic energy hopping is marked by $t$ and the antiferromagnetic Zeeman hopping is denoted by $t_{J}$. $\sigma_{0}$ and $\sigma_{z}$ are a unit matrix and a z-component Pauli matrix, respectively. Dotted lines mark $\mathcal{R}_{S}\mathcal{M}$ symmetries combining mirror plane with spin rotation.
\textbf{e} The spin d-wave interaction makes the electrons experience an opposite Zeeman field (magenta arrows) around the two magnetic valleys (TRIMs) $\boldsymbol{M}_{\text{1}}$ and $\boldsymbol{M}_{\text{2}}$ (calculated in the  ${\bf k}\cdot{\bf p}$ approximation, Eq.~2). We set $t=t_J/2$ in panel  \textbf{c} and  $t=0$ in panel \textbf{e}.}
\label{f1}
\end{figure}

\begin{figure}[h]
\hspace*{0cm}\epsfig{width=1\columnwidth,angle=0,file=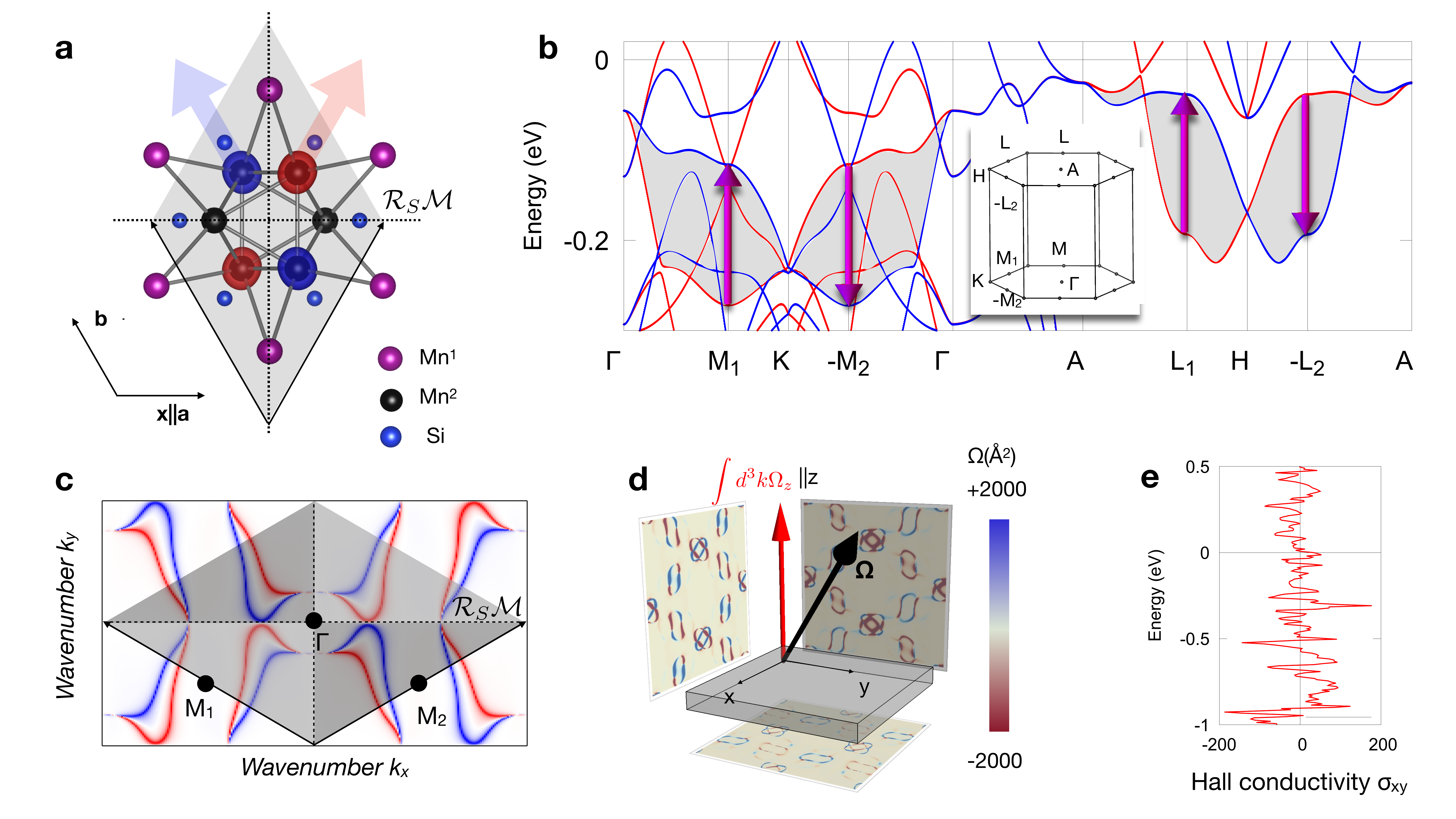}
\caption{\textbf{{\em Ab initio} theory of staggered spin-momentum interaction and spontaneous Hall effect in the collinear antiferromagnetic phase of Mn$_{\text{5}}$Si$_{\text{3}}$.} 
\textbf{a} Magnetization density isosurfaces of Mn$_{\text{5}}$Si$_{\text{3}}$ where red and blue colors indicate opposite spin polarizations. Dotted lines mark $\mathcal{R}_{S}\mathcal{M}$ symmetries.  \textbf{b} Calculated energy bands  split by the staggered spin-momentum interaction in the $k_{z}=0$ plane with relativistic spin-orbit interaction switched off. The staggered Zeeman field is marked by magenta arrows. 
\textbf{c}  $k_{z}=0$ Fermi surface cut calculated  without  spin-orbit coupling.
\textbf{d} Hall vector $\boldsymbol{\sigma}$ and crystal momentum resolved Berry curvature of band number 97 (band crossing Fermi level) have all three components non-zero. The panels with three Berry curvature projections show the same part of the Brillouin zone $k_{x},k_{y}\in (-2\pi,2\pi)$.  The N\'eel vector is set along the [111] direction and spin-orbit coupling is switched on. 
\textbf{e} Calculated anomalous Hall conductivity  $\sigma_{xy}$ corresponding to the component measured in the experiment.
 }
\label{f2}
\end{figure}

\begin{figure}[h]
\hspace*{0cm}\epsfig{width=1\columnwidth,angle=0,file=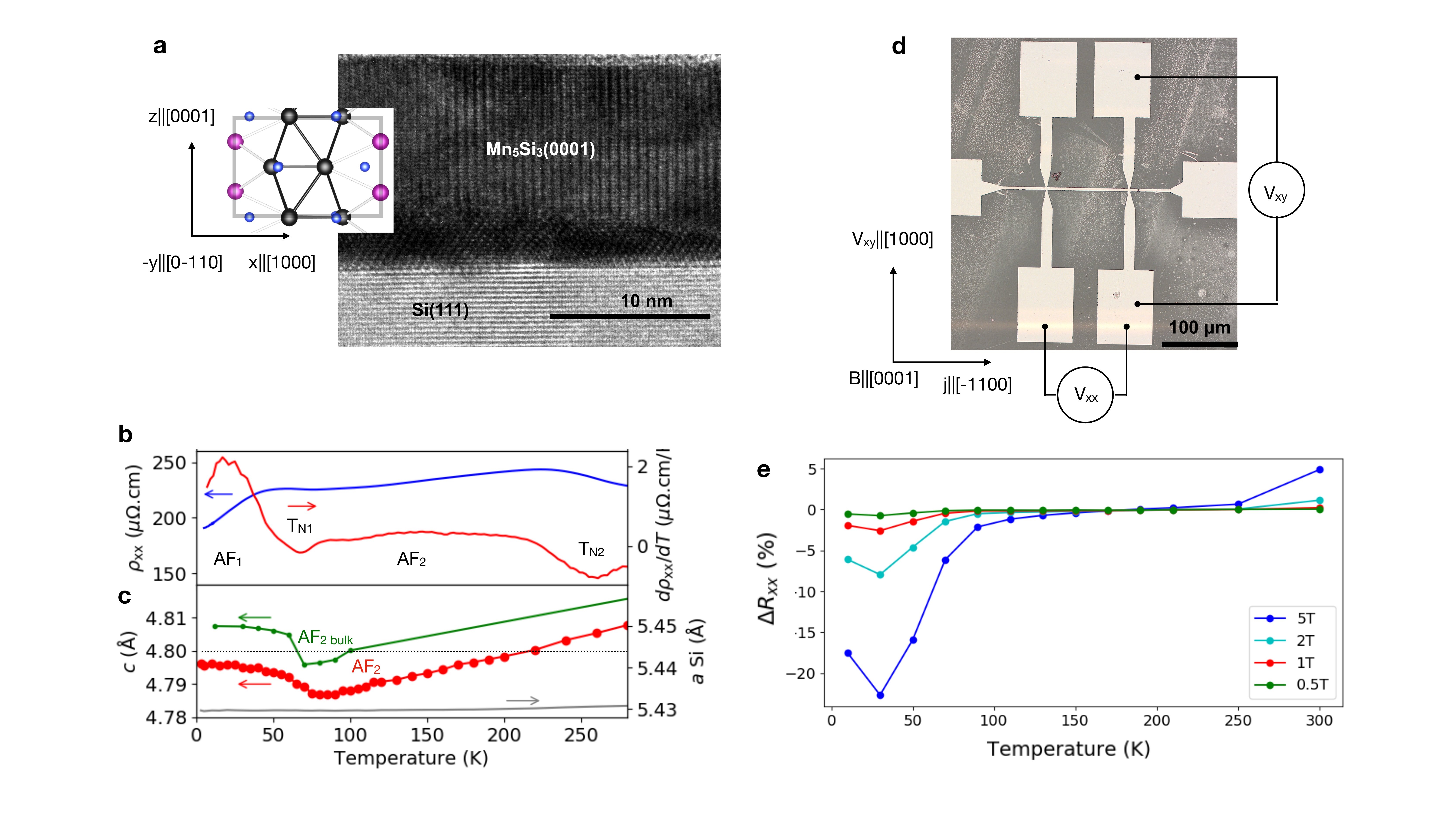}
\caption{\textbf{Structural and magnetotransport characterisation of Mn$_{\text{5}}$Si$_{\text{3}}$ epilayers.}
\textbf{a} Transmission electron microscopy image of a Mn$_{\text{5}}$Si$_{\text{3}}$ thin film grown on Si substrate. The inset shows the unit cell orientation confirmed via X-ray diffraction (XRD) characterization (see Methods and Supplementary Figs.~S5, S6).
\textbf{b} The temperature dependent longitudinal resistivity $\rho_{xx}$ and its derivative $d\rho_{xx}/dT$ reveal two antiferromagnetic phases AF$_{\text{1}}$ and AF$_{\text{2}}$.
\textbf{c} Lattice constant $c \parallel$~[0001] obtained from temperature dependent XRD compared to data for bulk Mn$_{\text{5}}$Si$_{\text{3}}$ taken from \cite{Gottschilch2012} (green).
\textbf{d} Optical micrograph of the litographically patterned Hall bar, orientation of the crystal, and applied magnetic field $\textbf{B}$.
\textbf{e} Temperature-dependent longitudinal magnetoresistance recorded with a magnetic field of 0.5,1, 2, and 5 T applied along [0001]. 
}
\label{f3}
\end{figure}

\begin{figure}[h]
\hspace*{0cm}\epsfig{width=1\columnwidth,angle=0,file=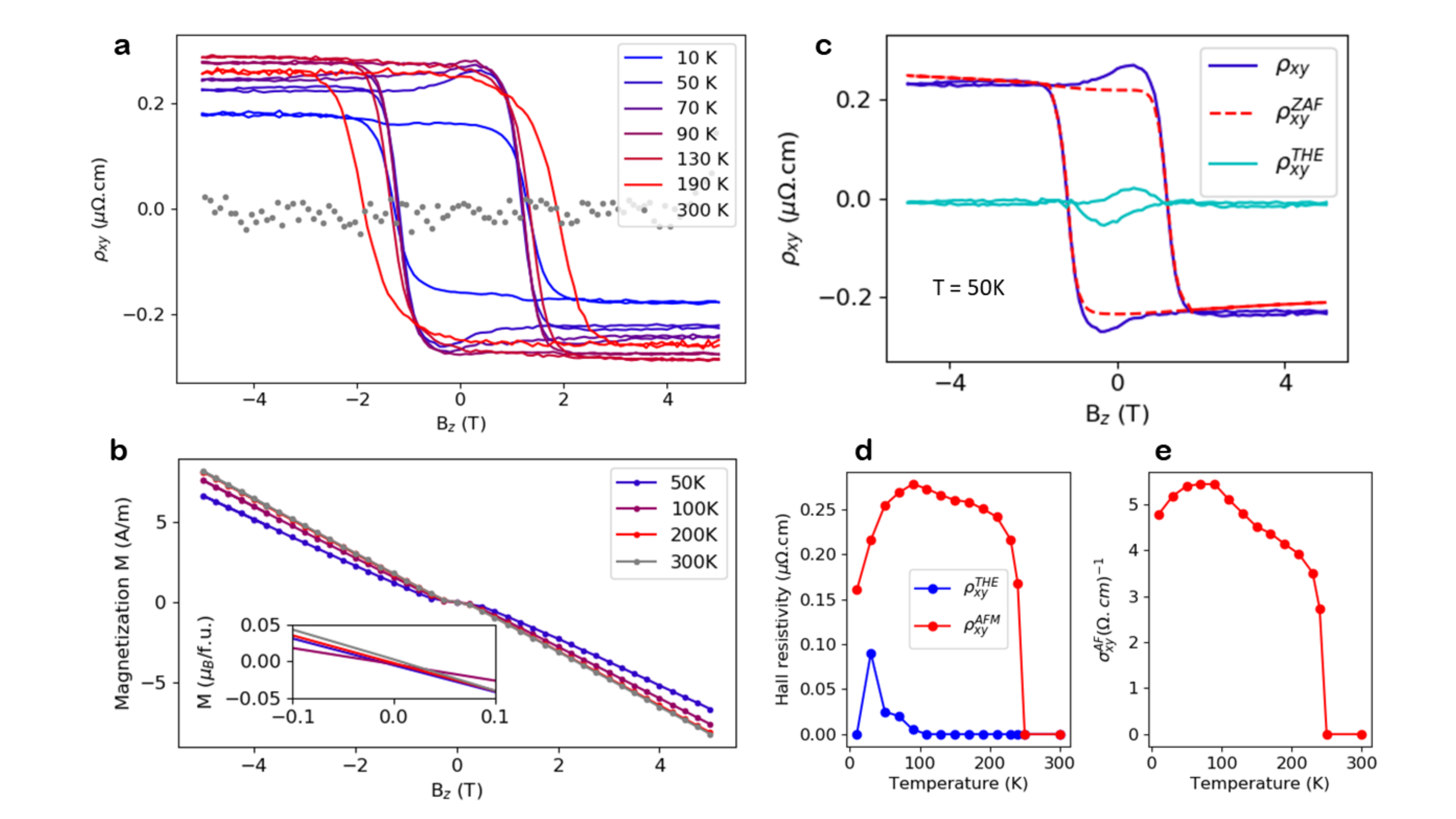}
\caption{\textbf{Experimental observation of  the anomalous Hall effect in Mn$_{\text{5}}$Si$_{\text{3}}$ epilayers.} 
\textbf{a} Anomalous Hall resistivity in the antiferromagnetic phase measured at several temperatures. The ordinary Hall effect, which is linear in the applied magnetic field $B_{z}$, was subtracted (see Methods and Fig.~S8). 
\textbf{b} Magnetization as a function of the magnetic field. The inset shows vanishing remanent magnetization and a small slope at low fields indicating the presence of compensated antiferromagnetism. 
\textbf{c} Decompositions of the Hall resistivity into a collinear $\rho_{xy}^{AFZ}$ and a noncollinear $\rho_{xy}^{THE}$ component at 50~K (see Methods and Supplementary Fig.~S8).
\textbf{d} Anomalous Hall resistivity as a function of temperature, decomposed into the collinear and noncollinear components.
\textbf{e} Anomalous Hall conductivity as a function of temperature. For Hall conductivities measured at another 4 samples see Supplementary Fig.~S9.} 
\label{f4}
\end{figure}
\end{document}